\documentstyle[12pt,fleqn,aaspp4]{article}
\newcommand{\be}{\begin{equation}}
\newcommand{\ee}{\end{equation}}

\begin{document}

\title{Anomalous RR Lyrae $(V-I)_0$ colors in Baade's Window}

\author{Amelia Stutz, Piotr Popowski\altaffilmark{1}\altaffiltext{1}{Ohio State 
University Presidential Fellow}, and Andrew Gould\altaffilmark{2}\altaffiltext{2}
{Alfred P.\ Sloan Foundation Fellow}}
\affil{Ohio State University, Department of Astronomy, Columbus, OH 43210}
\affil{E-mail: stutz,popowski,gould@astronomy.ohio-state.edu}

\begin{abstract}
We compare $(V-I)_0$--$(V-K)_0$ color-color and $(V-I)_0$--log $P$ 
period--color diagrams for Baade's Window and local RRab Lyrae stars.
We find that for a fixed log $P$ the Baade's Window RR Lyrae
stars are $\sim 0.17$ magnitudes redder in $(V-I)_0$ than 
the local RR Lyrae stars.  We also 
show that there is no such effect observed in $(V-K)_0$.  
We argue that an extinction misestimate towards Baade's Window is 
not a plausible explanation of the discrepancy. 
Unlike Baade's Window RR Lyrae stars, the local ones follow a black-body color--color 
relation and are well approximated by theoretical models.  
We test two parameters, metallicity and surface gravity, 
and find that their effects are too small to explain the $(V-I)_0$
discrepancy between the two groups of stars.
We do not provide any explanation for the anomalous $(V-I)_0$ 
behavior of the Baade's Window RR Lyrae stars.
We note that a similar effect for clump giant stars has been 
recently reported by Paczy\'nski and we caution that RR Lyrae stars 
and clump giants, often used as standard candles, can be subject to the 
same type of systematics. 
\end{abstract}

\section{Introduction}

RR Lyrae stars are used as standard candles to determine distances in the 
local Universe.  For a class of objects to be a 
standard candle, they must have a common luminosity, or else their 
luminosities must be a well-understood function of other observables such 
as period, metallicity, and environment.  
If one cannot determine the effects of such parameters and hence is not able to 
take them into account in the distance calculations, one cannot 
rely on the accuracy of the calculated distances.  
RR Lyrae stars are used to estimate distances to Baade's Window, 
globular clusters, and the Magellanic Clouds. 
An error in the calculation of the distance to the Large Magellanic Cloud (LMC) would 
carry over to the Hubble constant because the distance to the LMC is used as a 
standard for all Cepheid-calibrated secondary distance indicators.  
The inferred ages of globular clusters scale approximately as the inverse square
root of the distance and therefore inversely as the
luminosity of RR Lyrae stars.  As a result, if RR Lyrae stars fail to function as standard 
candles, there would be far-reaching consequences for our understanding of 
cosmological parameters.  

	Udalski (1998) measured the distance to the LMC using the RR Lyrae 
absolute magnitude--metallicity relation with the zero point calibrated by Gould \& 
Popowski (1998).  He obtained a distance modulus to the LMC of $18.09 \pm 0.16$. 
The difference between typical previous RR Lyrae distance determinations and the 
Udalski (1998) study is due to his large reddening correction.  Alcock et al. 
(1998) showed that the $(V-K)_0$--log $P$ relation for RR Lyrae stars provides a 
way to determine reddening.  We therefore hoped that the $(V-I)_0$--log $P$ 
relation could be used in a similar fashion to test Udalski's (1998) distance 
to the LMC.
However, every relation influencing the determination of the distance scale 
should first be proved to be environment-independent.
Hence, in \S2 and \S3 we compare the $(V-I)_0$--$(V-K)_0$ color-color and the $(V-I)_0$--log $P$  
relations for RR Lyrae stars from Baade's Window and the solar neighborhood.  We show 
that the local stars behave like black bodies, while the Baade's 
Window stars seem to have a different, non-black-body
spectral distribution.  We investigate two possible parameters that 
might drive this discrepancy, namely metallicity and surface gravity, 
but find no compelling case for either of them.  We present suggestive evidence that it is the 
$(V-I)_0$ rather than the $(V-K)_0$ colors of Baade's Window RR Lyrae stars that are anomalous.
In \S 4 we speculate on the significance of these results for the local distance scale.

\section{$(V-I)_0$ versus $(V-K)_0$ relation}

	Figure 1 is a dereddened color--color diagram which 
compares the $(V-I)_0$--$(V-K)_0$ colors of local RR Lyrae stars ({\it squares}) 
to those in Baade's Window ({\it circles}).
The $V$ and $I$ magnitudes for Baade's Window are based on Optical
Gravitational Lensing Experiment (OGLE) data (Udalski 1998).  
Udalski (1998) determined the extinctions from the Stanek (1996) 
map with the zero point corrected according to Gould, Popowski \& Terndrup (1998) 
and Alcock et al. (1998).  The $K$ magnitudes are taken from 
Carney et al. (1995).

	The data for the local RR Lyrae stars come from 
Jones et al.\ (1992) and from their sources (Liu \& Janes 1990; Jones, Carney, 
\& Latham 1988).  The original extinction corrections 
for the local RR Lyrae stars are described 
in Liu \& Janes (1990) and Jones et al.\ (1988).  As Gould \& Popowski (1998) argue,
the Schlegel, Finkbeiner, \& Davis (1998) extinction map is likely to become
a standard in Galactic and extragalactic astronomy. 
Therefore, we apply the following procedure to obtain dereddened colors. 
First we take the Jones et al.\ (1992) dereddened colors and find the 
observed colors using the $E(B-V)$ adopted by the authors.  We then deredden 
the colors using $E(B-V)$ from Schlegel et al.\ (1998) under the 
assumption that $R_V=A_V/E(B-V) = 3.1$.
The correction to the Jones et al.\ (1992) dereddened colors resulting from this 
procedure is very small ($\sim 0.01$ mag for an individual star). 
The Johnson $V$ and Cousins $I$ band passes are common 
to all studies.  All of the $K$ data are from the Carney group and so 
should be on the same system.

	There is a striking difference between the Baade's 
Window and local RR Lyrae stars. 
Since RR Lyrae stars are rather hot ($T_{\rm eff} \sim 6000$ K) and 
metal-poor, one expects them to behave approximately like blackbodies.  
In Figure 1 we therefore superpose a theoretical 
blackbody color-color curve on the stellar data ({\it solid line}).  The local stars
are well approximated by this curve while the Baade's Window stars lie 
$\sim 0.2$ magnitudes above it.   
One might suspect that this effect is due to 
incorrect extinction estimates towards Baade's Window. 
However, the reddening vector shown in Figure 1 as a dashed line 
is almost parallel to the tracks defined by the 
stars and is perpendicular to the offset between the two groups.  
Adjusting the extinction would therefore not remove the discrepancy. 

	Another possible explanation for the offset could be a 
difference in metallicity and/or surface gravity in the Baade's Window 
RR Lyrae stars as compared to the local stars.
To explore this possibility we consider theoretical models obtained from
P.\ Hauschildt (1998, private communication, courtesy of M.\ Pinsonneault; 
See also Hauschildt, Allard, \& Baron 1998).  
In Figure 2 we display $(V-I)_0$ vs. $(V-K)_0$ for different solar scaled
metallicities at a fixed temperature of $6000$ K and a fixed 
log gravity of $3.5$.  This figure
demonstrates that even in the generous metallicity range 
$-2.0 \leq {\rm [M/H]} \leq 0.0$ color differences 
are too small to account for the discrepancy 
seen in Figure 1. Nevertheless, we note that $(V-I)_0$ sensitivity to 
metallicity
increases dramatically at  ${\rm [M/H]} \sim 0.0$. In principle it is 
therefore possible that super-solar metallicities of RR Lyrae 
stars in Baade's Window would resolve the problem.
However, Walker \& Terndrup (1991) have measured metallicities of   
41 RR Lyrae type ab stars in Baade's Window and find $<{\rm[Fe/H]}>\sim-1.0$. 
If Baade's Window stars have element mixture similar to that of the Sun, then
one must reject super-solar metallicity as a solution.   

	We also test the possibility that a difference in surface 
gravity could cause the offset seen in Figure 1. 
In Figure 3 we plot $(V-I)_0$ vs. $(V-K)_0$ for three different surface 
gravities: $3.5$, $4.0$, and $4.5$, 
(open triangles, open squares, and open circles respectively), 
at a fixed ${\rm[Fe/H]}= -1.0$ and for a wide range
of effective temperatures (Hauschildt et al.\ 1998).  In Figure 3a we 
superpose the Baade's Window RR Lyrae stars on top of the model predictions and 
in Figure 3b we do the same for the local RR Lyrae stars. The local 
RR Lyrae stars (Fig.\ 3b) lie slightly below the $\log g = 3.5$ line,
which is expected given their typical gravities of $\log g \approx 2.84$ (see below).
These plots show that different gravities produce only small shifts  in the 
color-color relation for main-sequence stars.
We note that the $(V-I)_0$ discrepancy is reduced only when the
surface gravities of Baade's Window RR Lyrae stars are bigger than the gravities
of the local RR Lyrae sample.
The local RR Lyrae stars are in the ``right place'' already. Hence, to explain the 
colors of RR Lyrae stars in Baade's Window would require that their gravities
be enormously large. Again, this solution is unsatisfactory.

	Unfortunately, the analyses described in the two previous paragraphs 
suffer from three major limitations.
First, we use models for main sequence stars and not horizontal branch stars.
Second, there may be some effects intrinsic to atmospheres of pulsating stars
that are not present in main sequence stars, although Jurcsik (1998) argues 
that pulsating stars have colors similar to static ones.
Third, the lowest gravity for the models available to us have $\log g = 3.5$.
This is somewhat more than predicted by the very tight (scatter of 0.004)
empirical relation (Jurcsik 1998) between 
the log of surface gravity and the log period of the RR Lyrae stars: 
\begin{equation}
\log g = 2.473 - 1.226 \log P \approx 2.84 \label{rrgrav}
\end{equation}
for a typical $\log P \approx  - 0.3$.  The result given by this relation agrees 
nicely with the more general log $g$ vs. log $P$ relation for radially pulsating stars 
given by Fernie (1995).  Nevertheless we think that our conclusions are reasonably secure, although 
they cannot be considered absolutely certain.

\section{Period--color relations}

	The difference between the local and Baade's Window RR Lyrae stars 
described in \S 2 may be
caused either by an anomalous $(V-K)_0$, an anomalous $(V-I)_0$,
or a combination of both. Based on period-color diagrams, we argue in this 
section that the most likely source of the offset seen in Figure 1 is 
that Baade's Window RR Lyrae stars are anomalously red in $(V-I)_0$.
First we discuss why we suspect that $(V-K)_0$ is 
the same in both environments. 
Then we calculate how much extinction towards Baade's Window
is expected if the $(V-I)_0$ color is taken at its face value.
We again stress that it is not possible to adjust the Baade's Window 
extinction (using a standard reddening law, $R_V=3.1$) in such a way as to 
produce a consistent picture in both
$(V-K)_0$ and $(V-I)_0$.

	In Figure 4 we plot $(V-K)_0$ against log $P$ 
for local ({\it solid squares}) and Baade's Window 
({\it open circles}) RR Lyrae stars. The agreement is excellent.  
If the  extinction for Baade's Window were based solely on 
the $(V-K)_0$--log $P$ relation for RR Lyrae stars 
(Alcock et al. 1998), then one would expect both 
groups of stars to overlap by definition. However, we note
that the extinction determination based on the much cooler K giant stars
(Gould et al.\ 1998) gives almost exactly the same result.
This independent confirmation of the zero point of the Stanek (1996)
map implies that the local and Baade's Window
stars can in fact be described by the same $(V-K)_0$ vs. log $P$ relation.

	In Figure 5 we plot $(V-I)_0$ against log $P$ 
for local ({\it solid squares}) and Baade's Window ({\it open circles}) 
RR Lyrae stars. This period-color relation suffers 
from the same type of discrepancy as seen in Figure 1.    

	We therefore check whether there are any other environments in which
RR Lyrae colors behave in a similarly anomalous way. We examine
three globular clusters with available $V$ and $I$ photometry: M68 (Walker 1994), 
IC 4499 (Walker \& Nemec) and NGC 1851 (Saviane et al. 1998), 
which have metallicities ${\rm [Fe/H]}= -2.1$, 
$-1.65$, and $-1.28$, respectively.  We plot $(V-I)_0$ against log 
period for M68 (Fig.\ 6a), IC 4499 (Fig.\ 6b), and NGC 1851 (Fig.\ 6c).  
On the three sub-plots we also include the local stars ({\it solid squares}).
The zero points and slopes of the period--color 
relations for the two more metal-poor clusters (M68 and IC 4499) are in 
a very good agreement with the data for local stars.  However,
NGC 1851 (${\rm [Fe/H]}= -1.28$) shows a $(V-I)_0$ offset with respect to the local 
stars in the same direction as the Baade's Window data. From this plot one might 
draw the conclusion that the offset 
in the Baade's Window period--color relation is caused by metallicity.
This is the second hint that metallicity may play a role in explaining 
anomalous $(V-I)_0$ colors for the Baade's Window RR Lyrae stars. 
Therefore, in Figure 7 we plot $\Delta (V-I)_0$ against {\rm [Fe/H]}, where
\begin{equation}
\Delta (V-I)_0 = (V-I)_{0,BW} - (V-I)_{0,MODEL}, \label{deltavmi}
\end{equation}
and where the {\rm [Fe/H]} for individual stars are taken from Walker 
\& Terndrup (1991).  Our $(V-I)_{0,MODEL}$ is a quadratic curve described by
$(V-I)_{0,MODEL} = (0.015\pm 0.011)(V-K)_0^2 + (0.366\pm 0.030)(V-K)_0 + (0.068\pm 0.019)$,
fitted to a model by 
Hauschildt et al.\ (1998) with ${\rm [Fe/H]}= -1.0$ and $\log g= 3.5$.
This plot shows no clear trend with metallicity.

	To quantify the level of the discrepancy between the $(V-K)_0$ 
and $(V-I)_0$ period-color relations, we find the correction
$\Delta E(V-I)$ required to force the Baade's Window stars to overlap the
local stars on the period-$(V-I)_0$ diagram (Figure 5). 
We assume that the slopes of period-color relations for both
groups of stars are the same and we fit for the slope and two 
zero points (one for Baade's Window and one for local stars) simultaneously.
Our solution is \\[-0.5cm]
%\begin{description}
\[ \!\!\!\!\!\!\!\!\!\!\! \left\{  \rule[-0.8cm]{0cm}{0.85cm} \right. \]
\vspace{-2.83cm}
\begin{eqnarray}
\makebox[-1ex]{}(V-I)_{0,BW}\makebox[2.8ex]{}&= &(0.44\pm 0.12) [ \log P + 0.283 ] + (0.62\pm 0.01)\\
\makebox[-1ex]{}(V-I)_{0,LOCAL}&= &(0.44\pm 0.12) [ \log P + 0.283 ] + (0.45\pm 0.02).
\end{eqnarray}
%\end{description}
%\begin{minipage}[t]{0.3in}
%\[ \left\{  \rule[-1.7cm]{0cm}{1cm} \right. \]
%\end{minipage} \ \
%\begin{minipage}[t]{6in}
%\begin{eqnarray}
%\makebox[-5.5ex]{}(V-I)_{0,BW}\makebox[2.8ex]{}&= &(0.44\pm 0.12) [ \log P + 0.283 ] + (0.62\pm 0.01)\\
%\makebox[-5.5ex]{}(V-I)_{0,LOCAL}&= &(0.44\pm 0.12) [ \log P + 0.283 ] + (0.45%\pm 0.02).
%\end{eqnarray}
%\end{minipage}
The correction to $(V-I)_0$ is the difference between the zero points 
for the two groups of stars,
\begin{equation}
\Delta E(V-I) = 0.17 \pm 0.02 .\label{deltaevi}
\end{equation}
It produces an adjustment to the zero point of the extinction,
\begin{equation}
\Delta A_V = 2.5 \Delta E(V-I)= 0.42 \pm 0.05.\label{deltaav}
\end{equation}
We note that the average extinction in Baade's Window, $\left< A_{V,f} \right> = \left< A_{V,i} +  \Delta A_V \right>= 1.79 \pm 0.07$ suggested
by this analysis is in good agreement
with the Walker \& Mack (1986) determination slightly adjusted by Gould et al.\ (1998). 
This analysis yields $A_V \approx 1.86$ based 
on $E(B-V)$ of RR Lyrae stars.  However, it is quite possible that 
both $(B-V)_0$ and $(V-I)_0$ are systematically different for the RR Lyrae 
stars in Baade's Window.  We believe that 
the evidence in favor of the $(V-K)$-based extinction,
and consequently normal behavior of the $(V-K)_0$ color, is rather strong.
Jurcsik (1998) finds no $(V-I)_0 > 0.6$ among 272 field, globular-cluster, 
and Sculptor RR Lyrae stars while many RR Lyrae stars in Baade's Window have $(V-I)_0>0.6$
(see Fig.\ 1).  This supports the argument 
that it is the $(V-I)_0$ colors of the Baade's Window stars that
are anomalous.  The fact that $(V-I)_0$ color is redder than expected 
means that the effective temperatures of Baade's Window RR Lyrae stars 
based on $(V-I)_0$ may be too low and that they should be corrected 
upwards by $\sim$700 K.  We note that  McNamara (1997) finds for local
RR Lyrae stars used for Baade-Wesselink analyses that temperatures
inferred from optical colors are about 200-300 K hotter than temperatures
inferred from $(V-K)_0$.  This effect may or may not be related 
to the anomalous colors of Baade's Window stars.

\section{Conclusions}
	RR Lyrae stars are very often used as calibrated candles in the local
distance determinations. They are among the main indicators of the distance
to Galactic globular clusters and therefore influence the estimates of
globular cluster ages and theories of galactic formation, and allow one to 
pose severe constraints on cosmological models.
RR Lyrae stars are also often used to determine the distance
to the LMC. Because most of the extragalactic distance scale is tied to the
LMC, all errors in its distance modulus propagate into determinations
of the Hubble constant.
Therefore it is important that we understand the characteristics
of RR Lyrae stars and are aware of possible systematics resulting from
variations in their properties between different systems.

	We have shown that the $(V-I)_0$ colors of the bulge RR Lyrae 
stars behave in an anomalous way, distinct from the $(V-I)_0$ colors of 
local stars. We have argued that neither gravity differences between 
the local and Baade's Window stars nor the iron abundance differences
can be responsible for such behavior.However, we cannot rule out the 
possibility that Baade's Window stars have very high abundances of 
$\alpha$-elements that substantially change their spectral energy distribution. 
This suspicion is additionally supported by the similar color effect observed 
for red clump stars in Baade's Window (Paczy\'nski 1998).
It seems that the properties of the bulge RR Lyrae stars and clump giants
are very similar. Therefore we caution that distance determinations
based on these two types of stars may be subject to the same type of systematic
errors. If these systematic effects appear in parallel in RR Lyrae stars and
clump giants, they are not likely to be revealed by the direct comparison
between distances based on these two groups of stars.
If two different distance indicators produce similar results, it usually 
suggests that systematics in both measurements were reduced to negligible
levels. Such a convergence of results is therefore particularly appreciated.
Udalski (1998) argues in favor of a very short distance scale to
the LMC using this very argument.  Our results argue for 
caution when deriving strong conclusions from such a convergence.

\acknowledgments 
 
	We would like to thank I. Saviane for sending us the 
periods of the RR Lyrae stars in NGC 1851.  
We are grateful to M. Pinsonneault for pointing us to the theoretical
models of Hauschildt et al. (1998).  
We are in debt to A. Udalski for explaining some 
technical details of the OGLE observations.
This work was supported in part by grant AST 97-27520 from the NSF.

\clearpage

\begin{figure}
\caption[junk]{\label{fig:one}
Color--color diagram 
comparing the $(V-I)_0$--$(V-K)_0$ colors of local RR Lyrae stars ({\it squares}) 
to those in Baade's Window ({\it circles}).  The solid line is the black body theoretical curve.
The dashed line represents an extinction vector.  Note that extinction could not correct the
discrepancy seen between the Baade's Window RR Lyrae stars and either the local stars
or the black body curve.
}
\end{figure}
 
\begin{figure}
\caption[junk]{\label{fig:two}
Color--color diagram comparing theoretical
$(V-I)_0$ and $(V-K)_0$ for different solar scaled
metallicities at a fixed temperature of $6000$ K and a fixed 
log gravity of $3.5$.  Note that even in the metallicity range of 
${\rm [M/H]}= -2.0$ to ${\rm [M/H]}= 0.0$ color differences 
are too small to account for the discrepancy 
seen in Fig.\ 1.
}
\end{figure}

\begin{figure}
\caption[junk]{\label{fig:three}
Color--color diagram of theoretical models (Hauschildt et al.\ 1998) 
comparing  $(V-I)_0$ vs. $(V-K)_0$ for three different surface gravities: $3.5$, 
$4.0$, and $4.5$ ({\it open triangles}, {\it open squares}, and {\it open circles}, 
respectively).  Note that the shift in color due to gravity is small.  
These theoretical models are compared with the Baade's Window RR Lyrae stars [panel (a)]  
and the local RR Lyrae stars [panel (b)]. 
Note that in panel (b) the local RR Lyrae stars lie where expected 
given their typical $\log g \approx 2.84$.  Panel (a) implies 
that the colors of the Baade's Window RR Lyrae stars could be explained 
by gravity effects only if their gravities were unacceptably large.
}
\end{figure}

\begin{figure}
\caption[junk]{\label{fig:four}
Period--color diagram which compares $(V-K)_0$ vs. log $P$ for local ({\it solid squares})
and Baade's Window ({\it open circles}) RR Lyrae stars.  
The agreement is excellent.
}
\end{figure}

\begin{figure}
\caption[junk]{\label{fig:five}
Period--color diagram comparing $(V-I)_0$ vs. log $P$ for local ({\it solid squares})
and Baade's Window ({\it open circles}) RR Lyrae stars.  
This figure shows the same type of discrepancy as seen in Fig.\ 1.
}
\end{figure}

\begin{figure}
\caption[junk]{\label{fig:six}
Period--color diagrams comparing $(V-I)_0$ vs. log $P$ for three globular 
cluster ({\it open circles}) and local ({\it solid squares}) RR Lyrae stars. 
Panel (a) presents the M68 RR Lyrae stars with metallicities 
${\rm [Fe/H]} = -2.1$.
Panel (b) shows the IC 4499 RR Lyrae stars with metallicities 
${\rm [Fe/H]} = -1.65$.
Panel (c) displays the NGC 1851 RR Lyrae stars with metallicities
${\rm [Fe/H]} = -1.28$.
The zero points and slopes of the two metal poor 
clusters (M68 and IC 4499) are in good agreement with the data for local stars.
However, the more metal rich cluster, NGC 1851, shows a $(V-I)_0$ offset
with respect to the local stars in the same direction as the Baade's Window data.
}
\end{figure}

\begin{figure}
\caption[junk]{\label{fig:seven}
The difference, $(V-I)_{0,BW} - (V-I)_{0,MODEL}$, vs. the Walker \& Terndrup (1991)
metallicity {\rm [Fe/H]} for individual RR Lyrae stars in Baade's Window.
Our $(V-I)_{0,MODEL}$ is based on the Hauschildt et al.\ 
(1998) theoretical model for ${\rm [M/H]}= -1.0$
and $\log g= 3.5$.  This plot shows no clear
trend with the {\rm [Fe/H]} metallicity of RR Lyrae stars.
}
\end{figure}


\clearpage

\begin{references}
\reference{alc} Alcock, C.\ et al.\ 1998, \apj, 494, 396
\reference{car} Carney, B.W., Fulbright, J.P., Terndrup, D.M., Suntzeff, N.B., 
\& Walker, A.R.\ 1995, \aj, 110, 1674
\reference{fer} Fernie, J. D.\ 1995, \aj, 110, 2361
\reference{goup} Gould, A., \& Popowski, P.\ 1998 \apj, 508, 000 (astro-ph/9805176)
\reference{gou1} Gould, A., Popowski, P., \& Terndrup, D. T.\ 1998, \apj, 492, 778
\reference{haus} Hauschildt, P. H., Allard, F., \& Baron, E.\ 1998, preprint.\ (astro-ph/9807286)
\reference{JCSL1} Jones, R.V., Carney, B.W., \& Latham, D.W.\ 1988, \apj, 326, 312
\reference{JCSL2} Jones, R.V., Carney, B.W., Storm, J., \& Latham, D.W.\ 1992,
\apj, 386, 646
\reference{Jur} Jurcsik, J.\ 1998, \aap, 333, 571 
%\reference{lay} Layden, A.C., Hanson, R.B., Hawley, S.L., Klemola, A.R., \& Hanley, C.J.\ 1996, \aj, 112, 2110
\reference{LJ} Liu, T., \& Janes, K.A.\ 1990, \apj, 354, 273
\reference{mcna} McNamara, D.H.\ 1997, \pasp, 109, 1221
\reference{pacz} Paczy\'nski, B.\ 1998, submitted to Acta Astron.\ (astro-ph/9807173 )
%\reference{pop} Popowski, P. \& Gould, A.\ 1998, \apj, 506, 000
\reference{sav} Saviane, I., Piotto, G., Fagotto, F., Zaggia, S., Capaccioli, M., Aparicio, A.\ 1998, \aap, 333, 479
\reference{Sch} Schlegel, D.J., Finkbeiner, D.P., Davis, M.\ 1998, \apj, 500, 525
\reference{sta} Stanek, K.Z.\ 1996, \apj, 460, L37 
\reference{uda} Udalski, A.\ 1998, Acta Astron., 48, 113
\reference{wal} Walker, A. R.\ 1994, \aj, 108, 555
\reference{wam} Walker, A. R., \& Mack, P.\ 1986, \mnras, 220, 69
\reference{wan} Walker, A. R., \& Nemec, J. M.\ 1996, \aj, 112, 202
\reference{wat} Walker, A. R., \& Terndrup, D. T.\ 1991, \apj, 378, 119
\end{references}
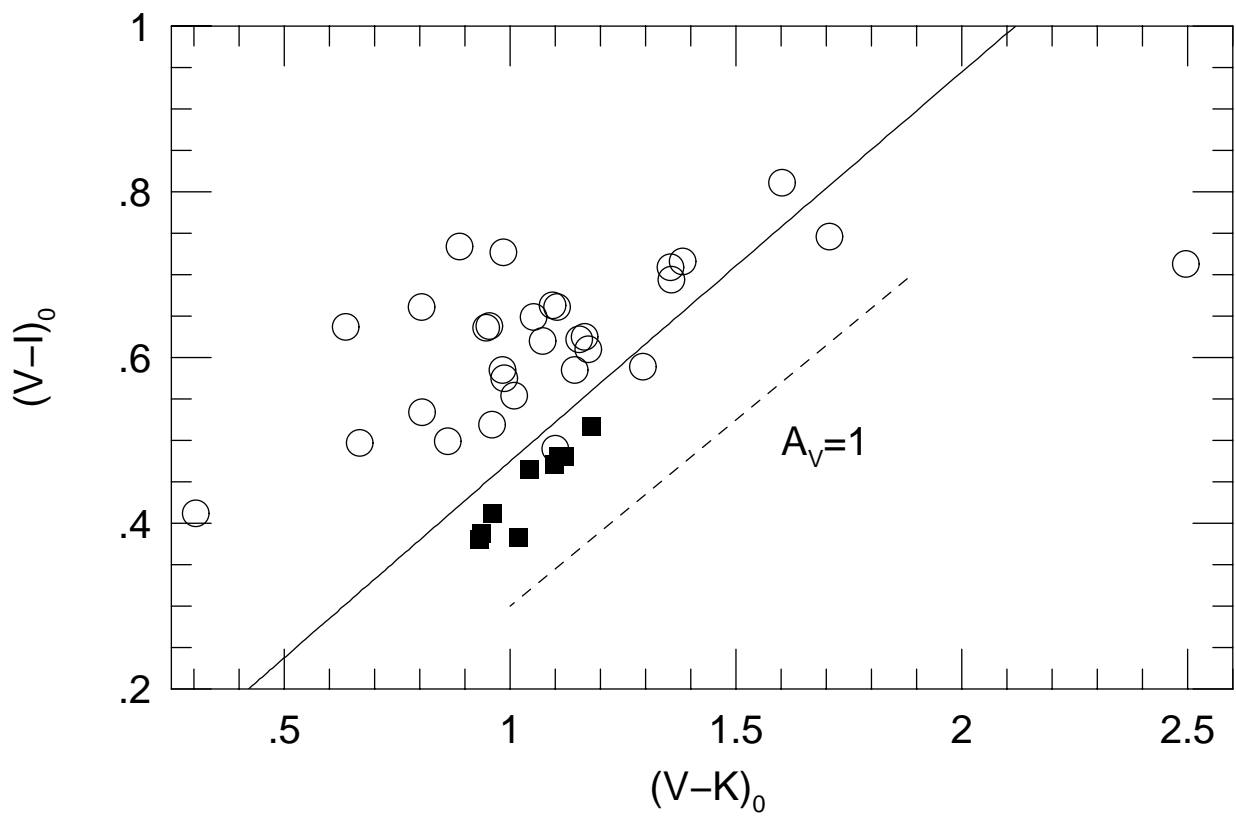
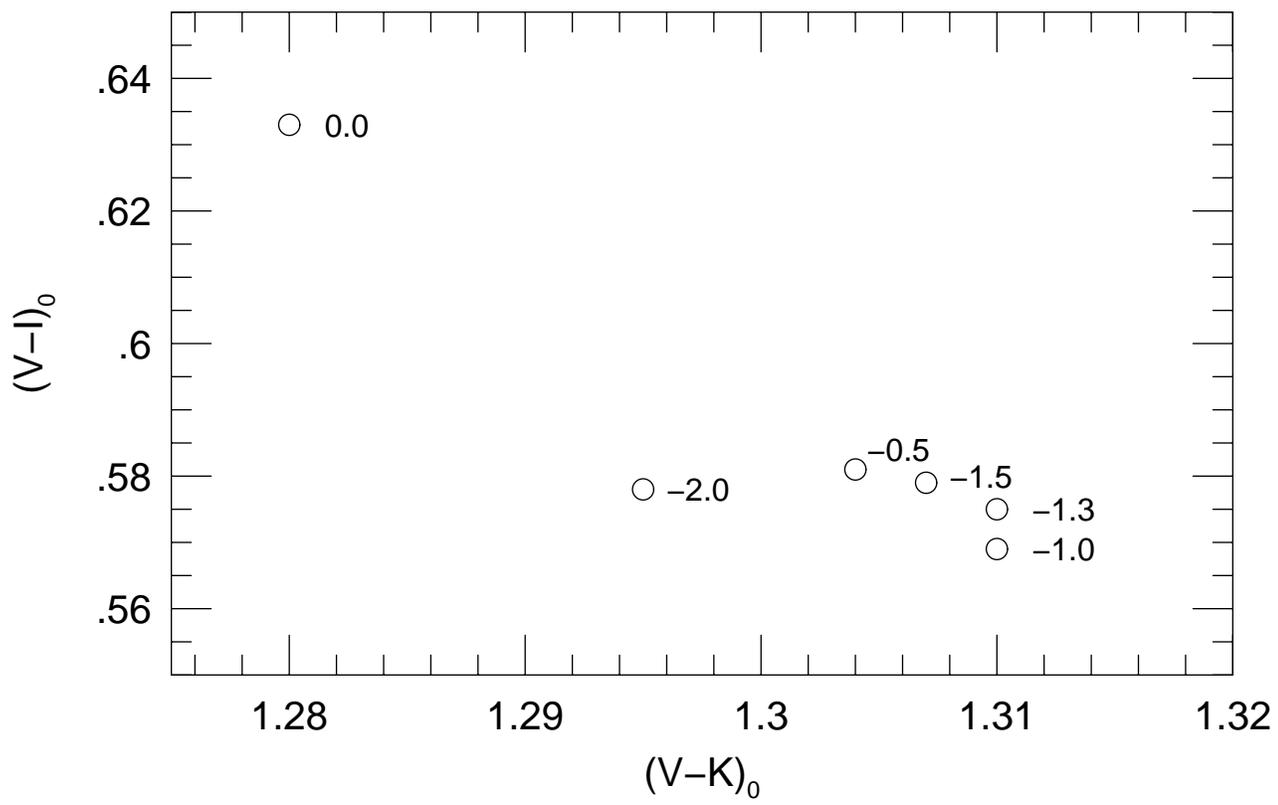
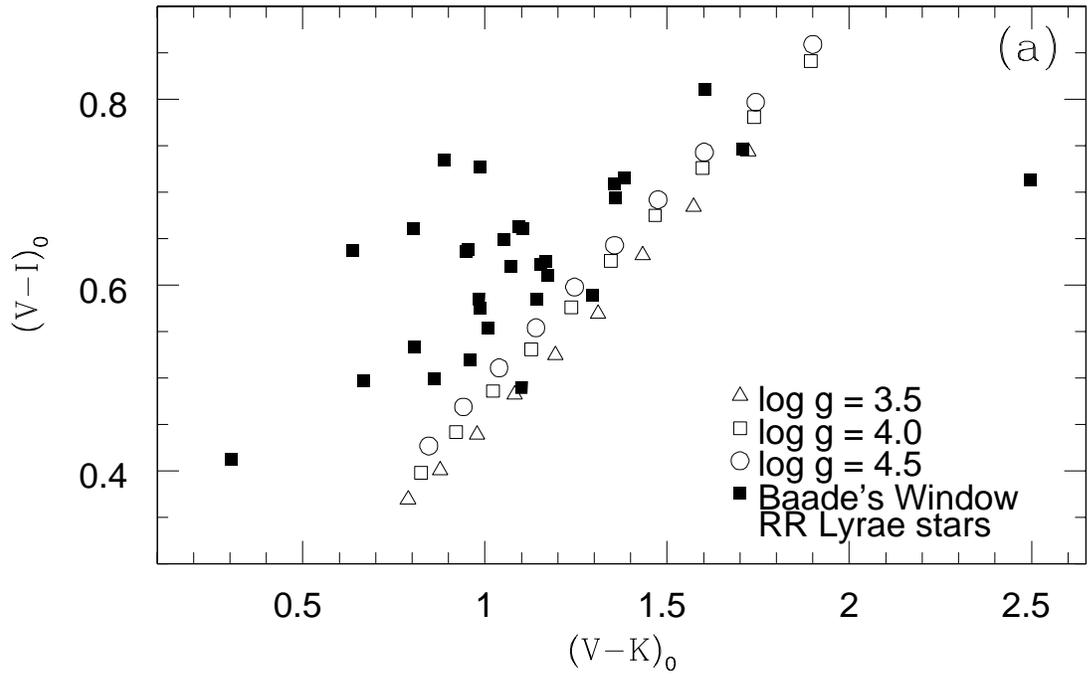
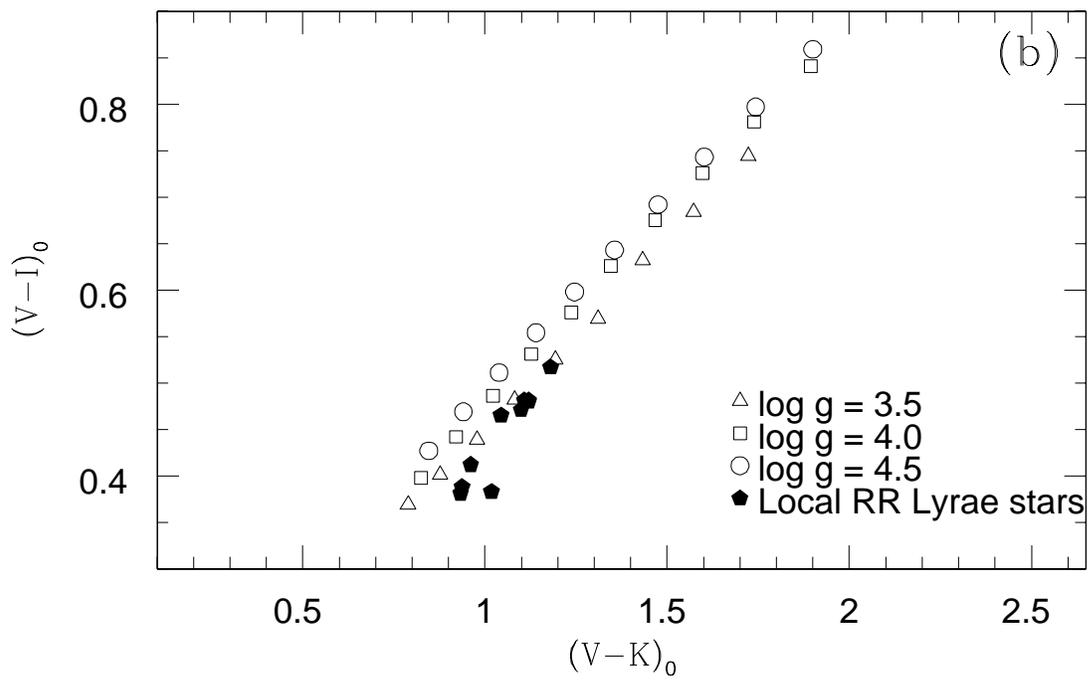
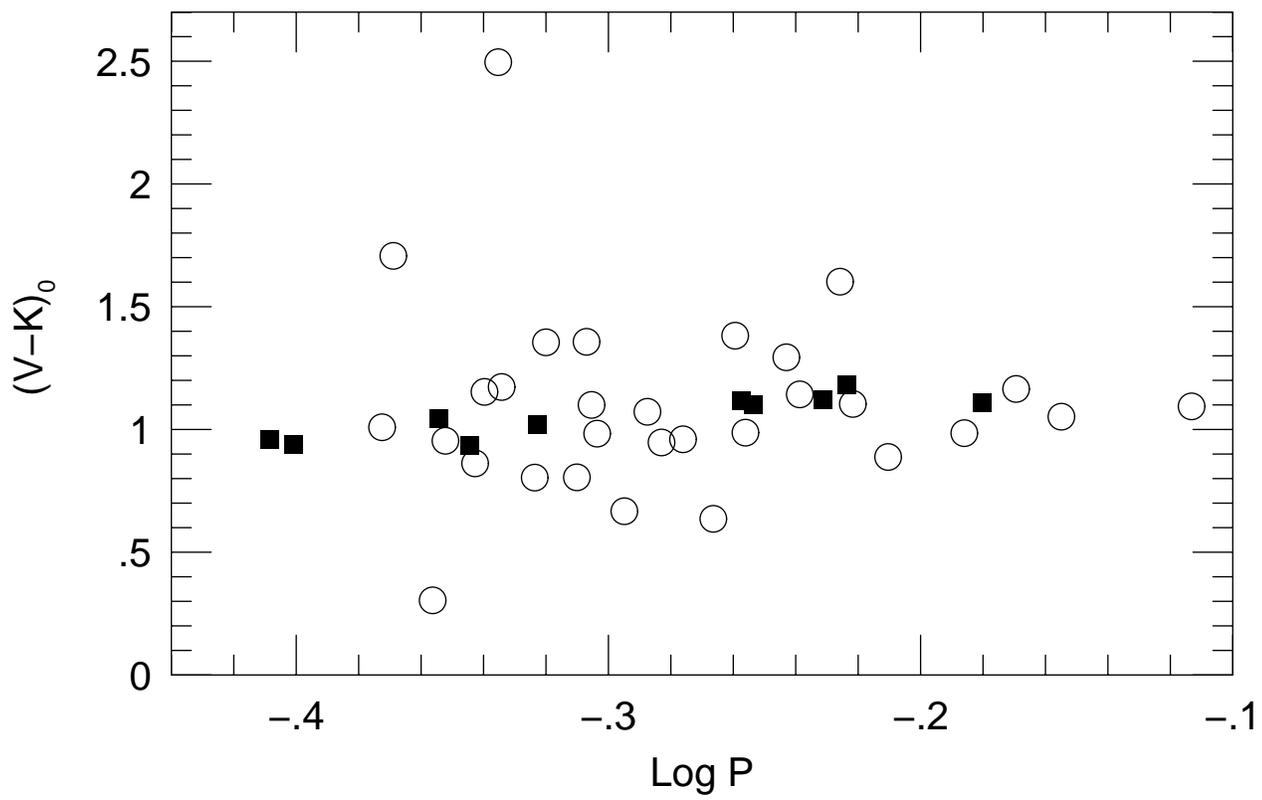
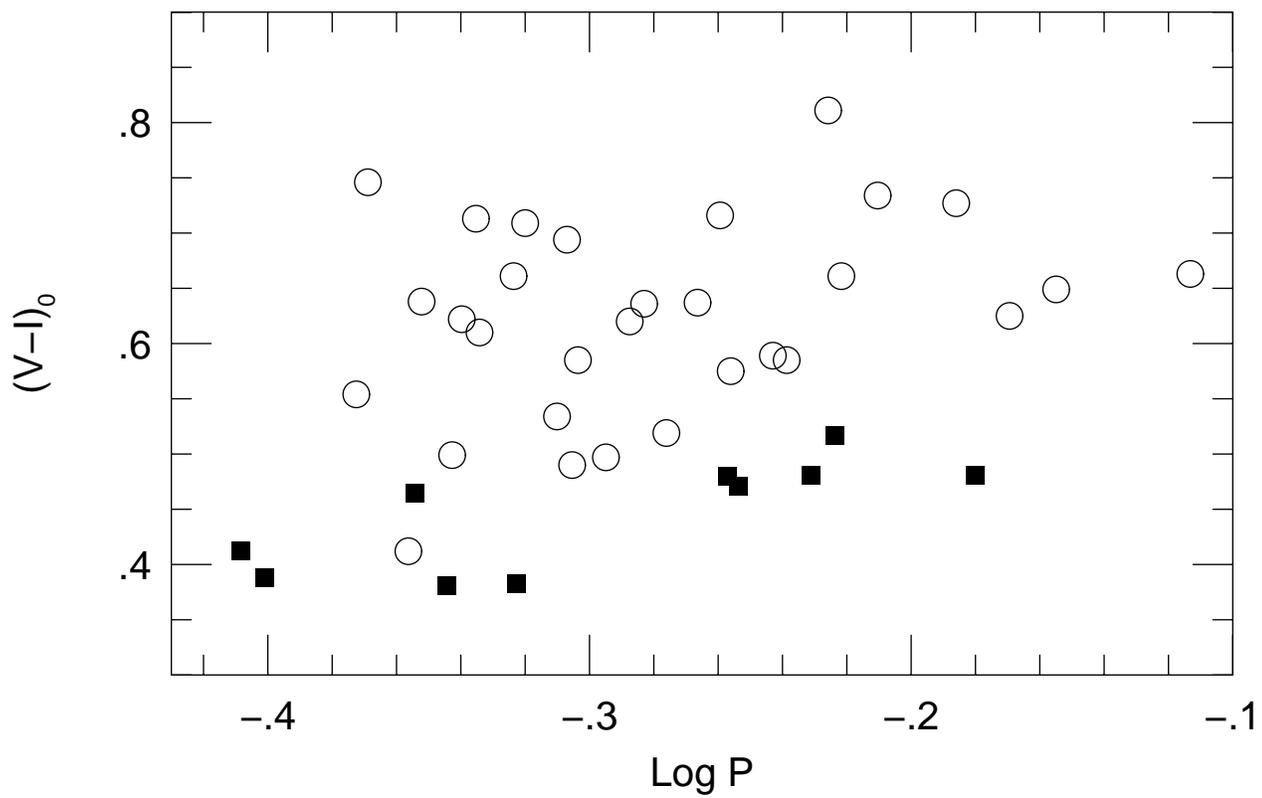
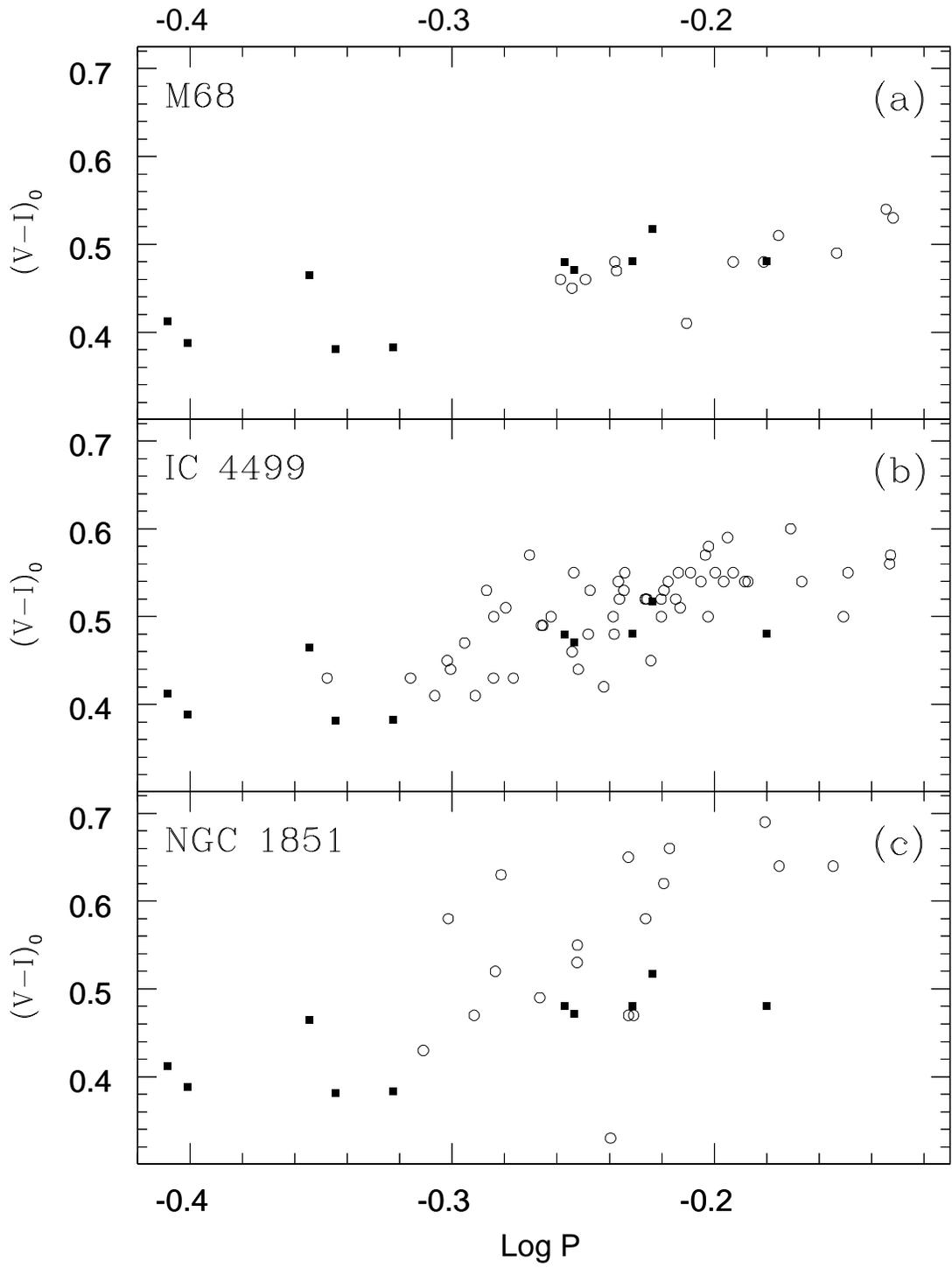
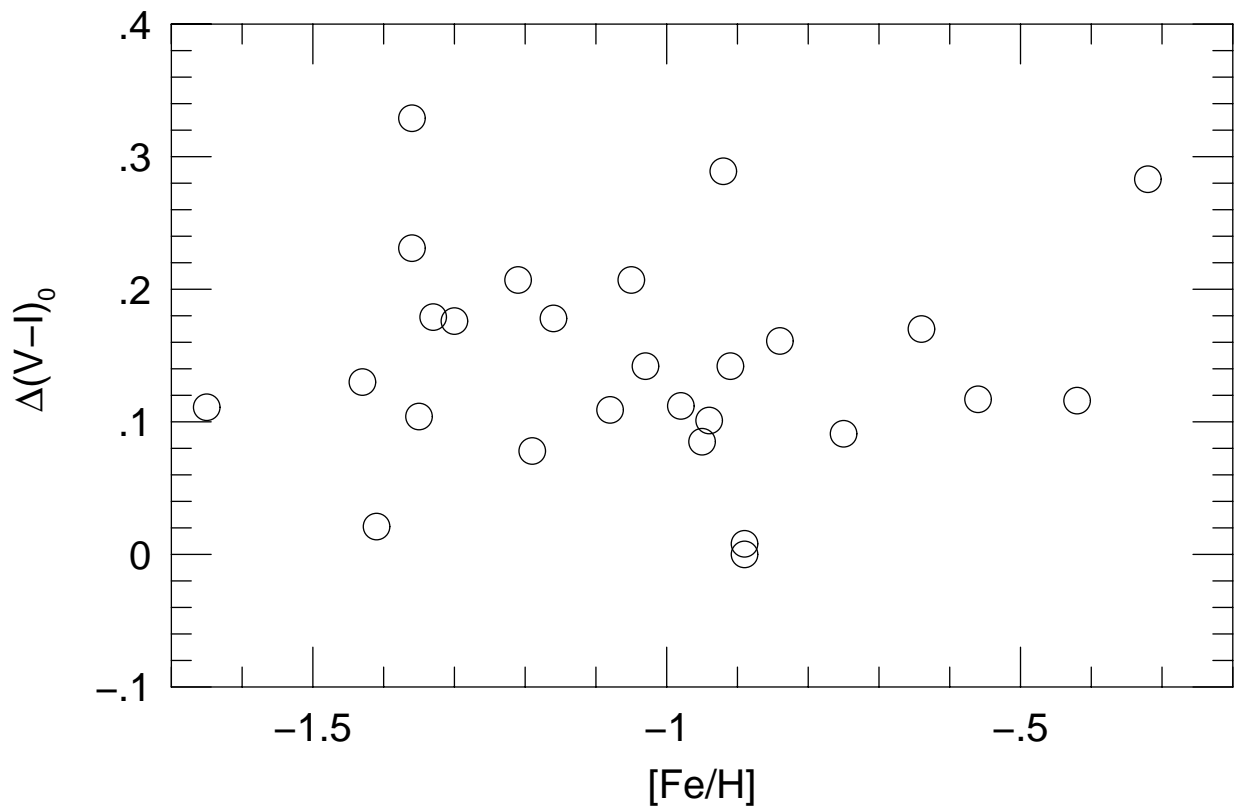
\end{document}